\documentclass[twoside,final]{pro12}

\usepackage[ansinew]{inputenc}
\usepackage{amsmath,amssymb,xcolor,journal_names}
\usepackage{graphicx}
\usepackage{array}
\usepackage{lscape}
\usepackage{multirow}
\usepackage{float}
\usepackage{natbib}
\usepackage{hyperref}
\newcommand{\MYhref}[3][blue]{\href{#2}{\color{#1}{#3}}}
\newcommand{\nup}{$\mathrm{\nu_p}$}
\newcommand{\N}{$\mathrm{n_e}$}
\newcommand{\dnn}{$\mathrm{\delta n_e/n_e}$}
\newcommand{\Rsun}{$\mathrm{R_\odot}$}
\newcommand{\Ms}{$\mathrm{M_{*}}$}

\newcommand{\Tbp}{T$_\mathrm{B}(\nu)$}

\newcommand{\Rx}{$\mathrm{R_X}$}
\newcommand{\Prot}{P$_\mathrm{rot}$}
\newcommand{\RHK}{R$^\prime_\mathrm{HK}$}
\newcommand{\alfmm}{$\alpha_{mm}$}
\newcommand{\teff}{T$_\mathrm{eff}$}

\usepackage{lineno}
\head{Mohan, A.} {Solar and stellar atmospheric tomography with modern telescopes}	% short title

\begin{document}

\title{Solar - stellar atmospheric tomography with mm-radio snapshot spectroscopic imaging}	% use capital letterss
\author{A. Mohan\adress{\textsl Rosseland Centre for Solar Physics, University of Oslo, Postboks 1029 Blindern, N-0315 Oslo, Norway}$\,\,^,$\adress{\textsl Institute of Theoretical Astrophysics, University of Oslo, Postboks 1029 Blindern, N-0315 Oslo, Norway}
}

\maketitle

\begin{abstract}
 Millimter (mm) frequencies are primarily sensitive to thermal emission from layers across the stellar chromosphere up to the transition region, while metrewave (radio) frequencies probe the coronal heights. Together the mm and radio band spectroscopic snapshot imaging enables the tomographic exploration of the active atmospheric layers of the cool main-sequence stars (spectral type: FGKM), including our Sun. Sensitive modern mm and radio interferometers let us explore solar/stellar activity covering a range of energy scales at sub-second and sub-MHz resolution over wide operational bandwidths. The superior uv-coverage of these instruments facilitate high dynamic range imaging, letting us explore the morphological evolution of even energetically weak events on the Sun at fine spectro-temporal cadence. This article will introduce the current advancements, the data analysis challenges and available tools. The impact of these tools and novel data in field of solar/stellar research will be summarised with future prospects.  
\end{abstract}

\section{Introduction}
\label{sec:intro}
Solar and stellar activity refers to a range of phenomena happening across the atmospheres of the sun/other cool main sequence stars (spectral type F - M), which lead to observable variability in their typical background/quiescent emission. These phenomena are driven by a variety of physical processes occurring across various outer atmospheric layers of the star starting from photosphere to corona \citep[see,][for an overview]{1997ApJ...487..437M,Cliver22_extremeSolarEvents_Rev}
Such active phenomena vary over a wide range of spatial, temporal and energy ranges. 
Their spatial scales vary within $\sim$ 100 - 10$^5$\,km, 
\citep[e.g.][]{2004JGRA..109.7105Y,Shibata07_reconn_spatial_scales,2022ApJ...926L...5N}, time scales from a few ms to hours\citep[e.g.][]{Osten2008,ash2012_flarestats,Hilaire13_typIIIstats,villadsen19_Cohbursts_but_notypeII, dal2020flare} and energy scales within $\sim$10$^{23}$ - 10$^{36}$\,erg \citep[e.g.][]{ash2012_flarestats,Namekata17_stats_solarflares_Vs_sunlikestars}.
Several of these phenomena finally impact the space-weather via large scale outflows and high energy particle activity, which can have huge impacts on the atmospheres of close-in planets\citep[e.g.][]{2010AsBio..10..751S, Vidotto13_highB_forclosebyplanets}. 
This makes the study of active phenomena on the sun (and stars) important from the perspective of their Earth/ionospheric-impact (and exo-planet habitability). 
In order to explore the cross atmospheric evolution of these phenomena,  given their variability scales, we need a high cadence ($\lesssim$\,1\,s) tomographic observing technique sensitive to the different heights across the active stellar atmospheric layers, namely chromosphere and corona.
It is within these active layers that different non-equilibrium processes dump a lot of excess magnetic field free energy to the local plasma resulting in frequent heating and particle acceleration over a range of energies~\citep[e.g.][]{Linsky16_Stellar_chromRev,ash2012_flarestats}.

\subsection{Tomographic exploration of active atmospheric layers}
Active atmospheric layers radiate energy across the entire electromagnetic spectrum, during both flaring and non-flaring periods.
The UV to soft X-ray continuum primarily tracks the coronal thermal emission from hot plasma, while the hard X-ray spectrum gives the information about accelerated particles.
These accelerated particles are expected to stream both upwards into interplanetary space generating radio bursts and downwards across chromospheric layers generating various spectral line emission signatures and finally photospheric white light flares~\citep{Namekata17_stats_solarflares_Vs_sunlikestars}.
Meanwhile, millimeter emission tracks chromospheric heating signatures.
%These accelerated particles are expected to stream both upwards into interplanetary space and also downwards into the atmosphere, causing shocks, heating and evaporation of lower atmospheric plasma triggering multi-scale flows~\citep{Namekata17_stats_solarflares_Vs_sunlikestars}. 
%The upward propagating high-energy electron beams set up the two-stream instability across overlying atmospheric layers, triggering a series of wave-wave and wave particle resonance interactions, leading to coherent radio bursts~\citep[see,][for an overview]{Melrose09_CoherentEmiss}. Occasionally, flares are associated with massive eruptions called coronal mass ejections (CMEs), which generate propagating shocks waves and subsequent acceleration of particles resulting in the type-II radio bursts~\citep[e.g][]{Leblanc01_CME-flare-typeII,vourlidas20_CME_review}.
%Meanwhile, photospheric heating by the supra-thermal plasma moving downwards from the acceleration region is often seen as prolonged white light flares lasting for a few minutes to several hours~\citep[e.g.][]{dal2020flare}.
\begin{figure}[!ht]
    \centering
    \includegraphics[width=0.72\textwidth]{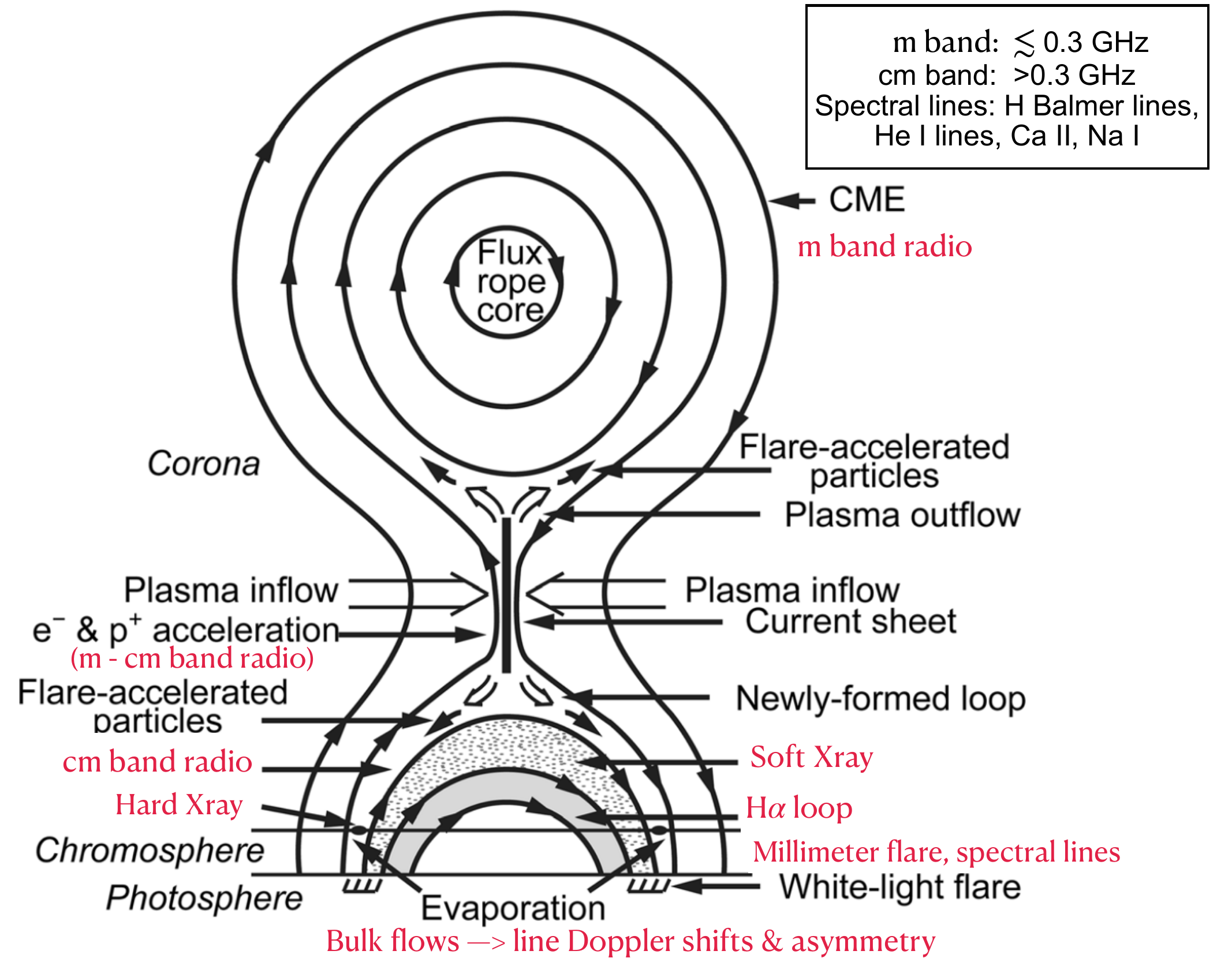}
    \vspace{-0.2cm}    
    \caption{Standard flare model showing a flaring active region loop. The magnetic reconnection site is shown with the current sheet and bi-directional accelerated particle beams marked. The different emission regions across all atmospheric layers are also shown (see, text for details). Important spectral lines are mentioned in the box. Evaporation and downflows produce spectral Doppler shifts and assymmetric line profiles. \citep[Figure credit:][]{Cliver22_extremeSolarEvents_Rev}.}
    \label{fig0_flaremodel}
\end{figure}
%The picture of the dynamics of matter and energy across the atmosphere will be incomplete without exploring the associated chromospheric impacts, since the chromosphere lies in between the photosphere and the corona.
%Spectral line diagnostics were the only means to infer chromospheric activity, until the dawn of sensitive millimeter (mm) interferometers like the Atacama Large Millimeter/sub-millimeter Array (ALMA)\footnote{\MYhref{https://www.eso.org/public/teles-instr/alma/}{https://www.eso.org/public/teles-instr/alma/}} and the Northern Extended Millimeter Array (NOEMA)\footnote{\MYhref{https://iram-institute.org/observatories/noema/}{https://iram-institute.org/observatories/noema/}}. 
%Millimeter emission is primarily formed at Local Thermodynamic Equilibrium (LTE), letting us track the thermal emission across the chromosphere~\citep[e.g.][]{Sven16_ALMA_science,MacGregor18_proxima_flares,2021RSPTA.37900185E, 2021RSPTA.37900174J}. 
See, Fig.~\ref{fig0_flaremodel} for an overview of the various emission sites during a typical solar flare.

\begin{figure}[t]
    \centering
    \includegraphics[width=\textwidth]{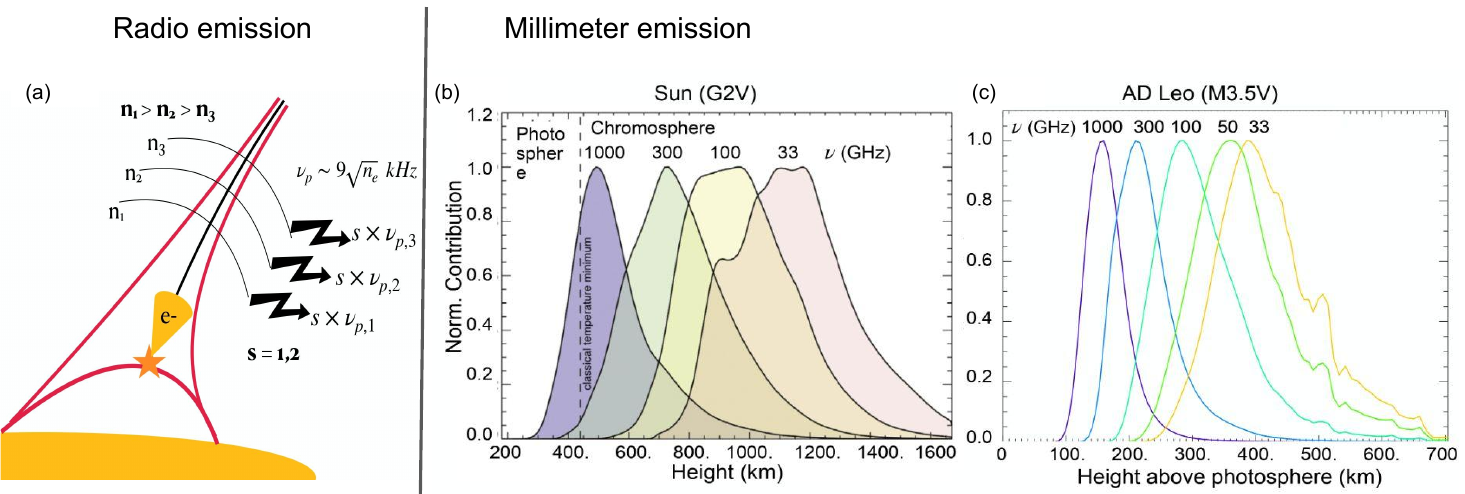}
    \vspace{-0.8cm}
    \caption{Graphical summary of the idea of tomographic imaging. (a)  Metrewave frequencies arising from varying atmospheric heights as the accelerated electron beams trigger instability across iso-density layers. (b) Millimeter emission contribution functions computed for the different ALMA bands using radiative transfer simulations applied on 3D atmospheric models. Plot for the sun is adapted from \citep{Sven16_ALMA_science}.}
    \label{fig1:tomography}
\vspace{-0.5cm}
\end{figure}

%Of the various wavebands, we need to find the wavelength ranges which can be used to perform tomographic exploration of the active atmospheric layers.
The Xray band though sensitive to corona, the emission can span a range of coronal heights since the waveband is optically thin. 
So, the Xray observations generally provide a line of sight averaged thermal and non-thermal evolution. 
However, in the case of the Sun, where spatially resolved imaging observations are possible, imaging of off disk-centre sources can be used to gain a sense for emission heights~\citep{1994Natur.371..495M}. 
But, this technique cannot be applied to stellar observations where sources are unresolved.
However, spectral line inversions can be done using simultaneous multi-line data to infer formation heights across chromosphere to corona. But, they suffer from various non-LTE and propagation effects
introducing degeneracy in the emission contribution function and the local physical mechanisms making it difficult to accurately infer emission heights~\citep[e.g.][]{2017A&A...605A..53M,2017ApJ...836...35J,2017PhRvL.118o5101K,2022ApJ...926..223Z}. 
Also, the time resolution obtained in multi-line spectral inversion studies is of the order of $\sim$ 10\,s to min even for stellar flares, making it difficult to trace seconds scale non-equilibrium dynamics, known to exist from high cadence radio/mm, UV and Xray observations~\citep[e.g.][]{Dennis85_HXR_highcadence_bursts,Osten2008,Endo10_HXRFlareCatalog,Hilaire13_typIIIstats}.

Meanwhile, meterwave radio bursts primarily originate as coherent emission across coronal heights, at the local plasma frequency (\nup)  and(or) its harmonic. Since local \nup\ is a function of local density ($\mathrm{n_e}$) in the corona, the emission heights can be inferred using a typical density model~\citep[e.g.][]{Reid2014,zucca2014}. 
This property lets us explore the propagation of high energy particles and active phenomena across coronal heights~\citep[e.g.][]{morosan2014,zucca2014,patrick2018_densmodel_frmtypIII}.
Figure~\ref{fig1:tomography}(a) graphically depict the idea of coronal tomography with radio imaging spectroscopy.
Similarly, emission at different mm frequencies form at LTE across chromospheric heights in sun-like stars with different frequencies having different penetration depths, acting as a linear thermometer~\citep{Sven16_ALMA_science}.
Figure~\ref{fig1:tomography}(b-c) presents the mm band contribution function models for the Sun and AD Leo (M4V).
However, in the case of strong M-dwarf flares from highly magnetised environments mm-continuum may track gyro-synchrotron emission. This can clearly be separated from the thermal contribution based on emission spectral index and polarisation evolution during the event, offering a unique means to get estimates for local magnetic fields and accelerated particle spectrum alongside thermal evolution across chromospheric layers~\citep[e.g.][]{MacGregor18_proxima_flares}.
Thus simultaneous mm and radio observations can provide a tomographic view of the dynamics and propagation of various active phenomena like flare, eruptions etc. across the active solar/stellar atmospheric layers from chromosphere to corona. Together with the high energy (Xray to optical) imaging spectroscopy, which provide complementary information on thermal and non-thermal plasma evolution across atmospheric layers, we can better constrain the models and infer local physics~\citep[e.g.][]{Osten2005_100percVpol,MacGregor18_proxima_flares, Zic20_typeIV_ProximaCen}.
Besides the active emission, quiescent mm emission helps probe the solar/stellar atmospheric heating gradient which is an important input to stellar atmospheric models~\citep[e.g.][]{2018MNRAS.481..217T,White20_MESAS,Atul21_EMISSAI}.
\paragraph{\textbf{Solar-stellar connection: }}Sun being the only spatially resolvable main sequence star, is a an excellent example to study the various types of radio/mm bursts (in time-frequency plane), their physical manifestations and space weather impacts using detailed image plane analysis. There has been several studies of solar flare and quiescent emission in radio and mm bands done in coordination with multi-waveband observations and modelling~\citep[see, ][for an overview]{SolarRadio_book, white2007_radiobursts, Sven16_ALMA_science}. These studies have provided valuable insights on the physics of various active phenomena and their multi-waveband observables. 
However, it is not necessarily right to directly extend the inferences from the numerous solar activity studies to stars.
For instance there has been no sign of type-II radio burst yet detected in the most active M-dwarf stars despite several long term monitoring campaigns~\citep[e.g.][]{villadsen19_Cohbursts_but_notypeII}. The number of CMEs itself is found to lower in active M dwrafs than what is expected from simply extending the solar paradigm~\citep[e.g.][]{Leitzinger14_nohalfCME, Odert20_CME_overview}.
A prominent belief is that, this is possibly due to the strong magnetic field strengths in M-dwarf coronae which results in high Alfv\'{e}n speeds, that in turn block most flares from causing eruptions. In case of erupting prominences, the high Alfv\'{e}n speeds could be stopping the shock formation and in turn causing the lack of type-II radio bursts which are associated with CME shocks\citep[e.g.][]{Mullan19_radioQuietCMEs_indMs, Odert20_CME_overview}.
Even young G dwarfs like EK\,Dra is known to deviate from standard solar flare picture atleast during superflares. Hours-long lasting white light flare with an extent similar to chromospheric H$\alpha$ flare has been reported in EK\,Dra~\citep{Namekata17_stats_solarflares_Vs_sunlikestars}. In solar flares, usually white light flares last for about half the duration of H$\alpha$ flares.
This anomalously long lived white light flaring in EK Dra is thought to be due to a possible radiative back-warming of photosphere from flare heated chromospheric plasma.
Besides, in the case of coherent metrewave ($\sim$\,100\,MHz) bursts, young active M dwarfs stars are known to produce highly polarised metric band flares often akin to Electron Cyclotron Maser emission (ECME), besides the coherent plasma emission mechanism~\citep[see,][for an overview]{vedantham21_mechCohEmissStars}. Around $\sim$\,100\,MHz, ECME mechanism is not that common in the sun given the magnetic field strengths and densities in the corona~\citep{gary2001}.
So while extending the solar paradigm, one should ensure that the stars are similar atleast in the sense of activity characteristics to the sun.
The F to early M type stars with mass greater than $\sim$0.35\,M$_{\odot}$ have an inner radiative and outer convective zone like the sun~\citep{chabrier00_Stellainterior}.
Besides, the stellar activity is known to evolve with age~\citep[e.g.][]{Skumanich72_Age-ro-activity, Donati09_Rev_Bfield,Vidotto14_B_Vs_age_n_rot}. 
\cite{Barnes03_Rot_Vs_age_Vs_Activity} showed that the stars older than 1\,Gyr typically falls in the low activity branch, usually referred to as `I'.
So the insights gained from solar observations could be extended to atleast sun-like stars belonging to F- early M spectral type and ages $\gtrsim$\,1\,Gyr. However, the plasma physics and emission mechanisms generally apply to all stars including the sun. So one needs to interpret the multi-waveband observations after incorporating the right local physical conditions like magnetic field strengths and atmospheric physical structure for the star of interest~\citep[see, ][for an overview]{guedel02_stellarradio,white04_solar-stellar_conn}.

With the advent of modern interferometric arrays like the MWA, LOFAR, uGMRT, MeerKAT, ALMA, NOEMA etc., sensitive high fidelity spectroscopic imaging observations (spectral resolution: $\sim$\,10\,kHz in radio; $\sim$ 2\,GHz (continuum) and 0.01 - 10\,MHz (spectral line) in mm band\footnote{\MYhref{https://almascience.eso.org/about-alma/alma-basics}{https://almascience.eso.org/about-alma/alma-basics}}) are now possible for the sun and nearby stars at $\lesssim$\,1\,s resolution. 
These facilities together cover $\sim$\,10\,MHz - 1\,THz wide band in a nearly seemless manner, enabling atmospheric tomography across chromosphere to outer corona (heliocentric height $\sim$ 2\,\Rsun) at high sensitivity~\citep[e.g.][]{white04_solar-stellar_conn,morosan2014,Sven16_ALMA_science,2020A&A...639L...7V,Atul21_EMISSAI}
%Thus, radio and millimeter continuum offers a unique means to perform tomographic exploration of solar/stellar activity at $\lesssim$\,1\,s owing to their emission mechanism and the supreme sensitivity offered by modern instrumentation~\citep[e.g.][]{white04_solar-stellar_conn,Sven16_ALMA_science,2020A&A...639L...7V,Atul21_EMISSAI}.
However, as always massive advancements comes with new challenges.
In this case, it is in the form of big data handling, reduction, analysis, automation and storage.

%Despite years of flare observations across various spectral bands, we still lack a clear physical understanding of the physical processes driving particle acceleration and heating across the active solar/stellar atmospheres in different flaring scenarios. Even in the case of the sun, where we do have a better understanding compared to stellar flares due to the large volume of spatially resolved data, we still can neither predict flares and post-flare phenomena like eruptions/outflows accurately, nor can we provide definitive models for different types of active phenomena.
%Much of this owes to the multiple degenerate possible mechanisms which are often hard to constrain with available data, lack of multi-waveband  spectroscopic information across Xray to radio waveband and a means to tomographicaly explore solar/stellar atmosphere during an active phenomenon at high spatio-temporal cadence and sensitivity. 
%For instance, most of the studies in both solar and stellar flares lack mm - radio co-observations. 
%In mm band we lacked sensitive instruments until recently, as mentioned earlier, while the previous generation radio imaging instruments often lacked simultaneous wideband spectroscopic snapshot imaging capability which is essential to study the propagating shocks, high energy particles and disturbances across coronal layers.

\section{Big data problem with modern radio/mm interferometers}

The high sensitivity and imaging fidelity of the modern instruments primarily owe to their compact large-N (N: antenna count) architecture, which enables dense u-v coverage. The high spectro-temporal sampling across a large-N array leads to high data rates exceeding $\sim$1\,TB/h. After analysis the data can expand further. Hence, automated calibration and imaging pipelines~\citep[e.g.][]{Atul17,Mondal19_AIRCARS} have been and are being built for modern telescopes alongside efficient flagging routines~\citep[e.g.][]{2010ascl.soft10017O,2023MNRAS.tmp..484Z}.
The final products of the calibration/imaging pipelines are usually calibrated measurement sets (MS) or 4-D image data cubes across time, frequency and angular sky-coordinates~\citep{Atul17}.
I will present two powerful tools which have been developed to analyse these final products and derive the flux evolution of the sources of interest in the image/visibility plane across fine spectro-temporal scales.

\subsection{SPatially REsolved Dynamic Spectrum (SPREDS)}
SPREDS is a tool intended to operate on calibrated 4D image data cubes of the Sun across spectro-temporal axes~\citep[see,][]{Atul17,Mohan2021b}.
SPREDS is a python code which can operate on across snapshot spectroscopic images parallely and extract flux density from regions of interest on the solar surface. The routine either records the source flux in a user-specified region of any geometric shape or attempts to fit a 2D Gaussian function to the morphology in the specified region to extract the evolution of the size, flux density and orientation of the source.
In cases where there is a bright source moving within some region in the image plane, the code can be customised to follow the source and fit 2D Gaussian functions to it. 

\begin{figure}[t]
    \centering
    \includegraphics[width=0.96\textwidth,height=0.34\textheight]{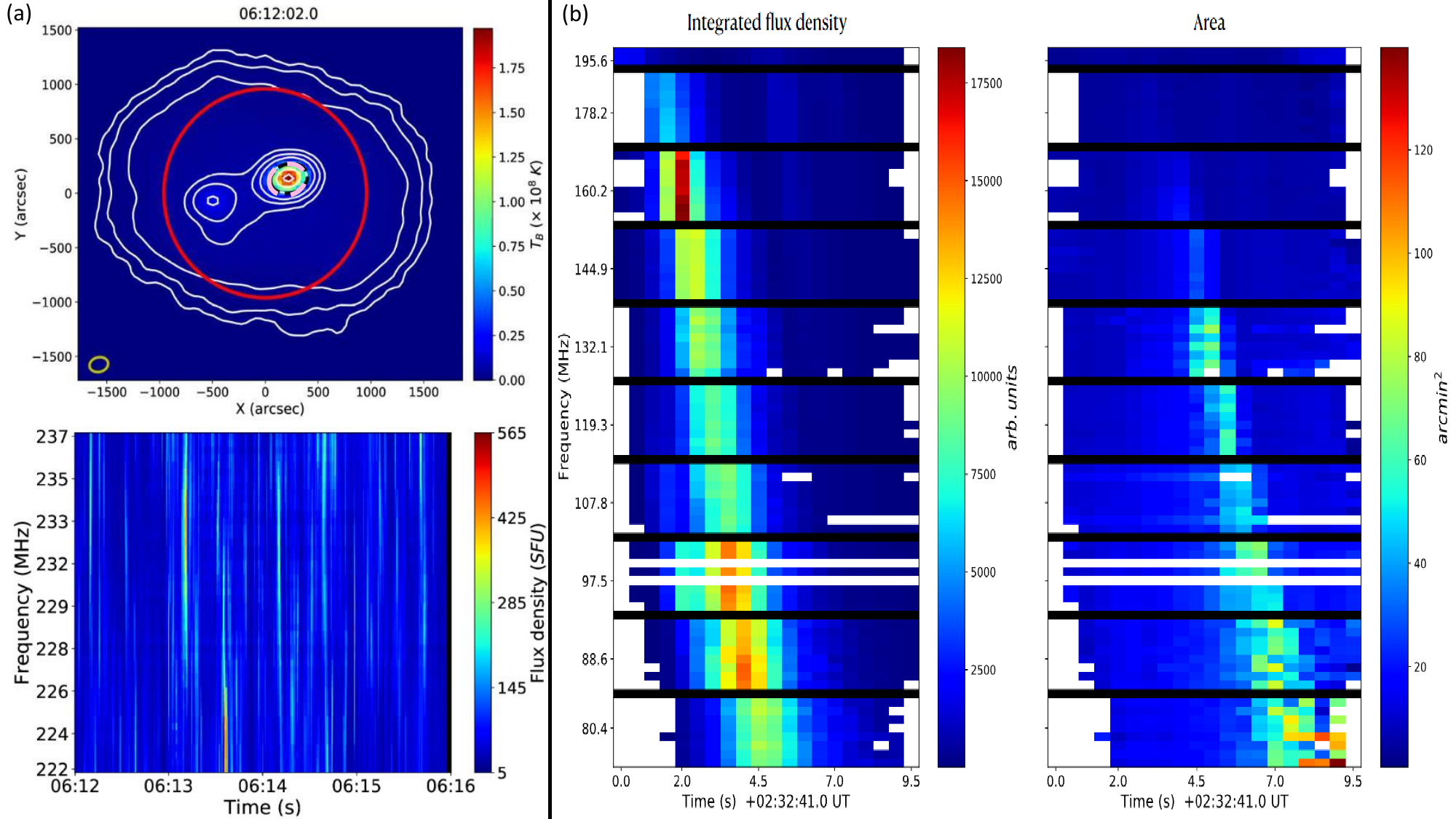}
    \vspace{-0.5cm}
    \caption{(a) \emph{Top}: A sample snapshot spectroscopic image made at 0.5\,s and 160\,kHz resolution using an MWA dataset from 2014-11-03~\citep{Atul19_ARTB_microflare}. Red ring marks the optical solar disk and the pink dotted ellipse marks the chosen 2$\times$psf region for deriving SPREDS shown in the \emph{Bottom} panel. (b): SPREDS derived by fitting 2D Gaussian functions to a type-III bust source in a different dataset~\citep{Atul21_dNN_vsht}.}
    \label{fig2:SPREDS}
\vspace{-0.2cm}
\end{figure}

Figure~\ref{fig2:SPREDS}(a) shows an example SPREDS for an elliptical fixed region of the size twice the synthesised beam. 
The snapshot spectroscopic image (resolution: 0.5\,s ; 160\,kHz) in the top panel shows the chosen fixed region in black dotted line. Red circle in the image marks the solar disk as seen in visible light data and the synthesised beam (psf) is shown in the bottom left corner of the image. 
Meanwhile, Figure~\ref{fig2:SPREDS}(b) shows the SPREDS derived by fitting 2D Gaussian functions to another burst source close to disk centre, from a different dataset. The resultant SPREDS based on the fitting procedure shows the spectro-temporal evolution of both source area (reported as the elliptical cross sectional area of the best-fit 2D-Gaussian function) and its integrated flux density. 
A working version of the code is made public via Github\footnote{\MYhref{https://github.com/atul3790/SPREDS}{https://github.com/atul3790/SPREDS}}, though with less documentation. A more user friendly version with detailed explanation will be released soon. 

\subsection{VISibility Averaged Dynamic spectrum (VISAD)}
\label{sec:VISAD}
\begin{figure}[t]
    \centering
    \includegraphics[width=\textwidth,height=0.25\textheight]{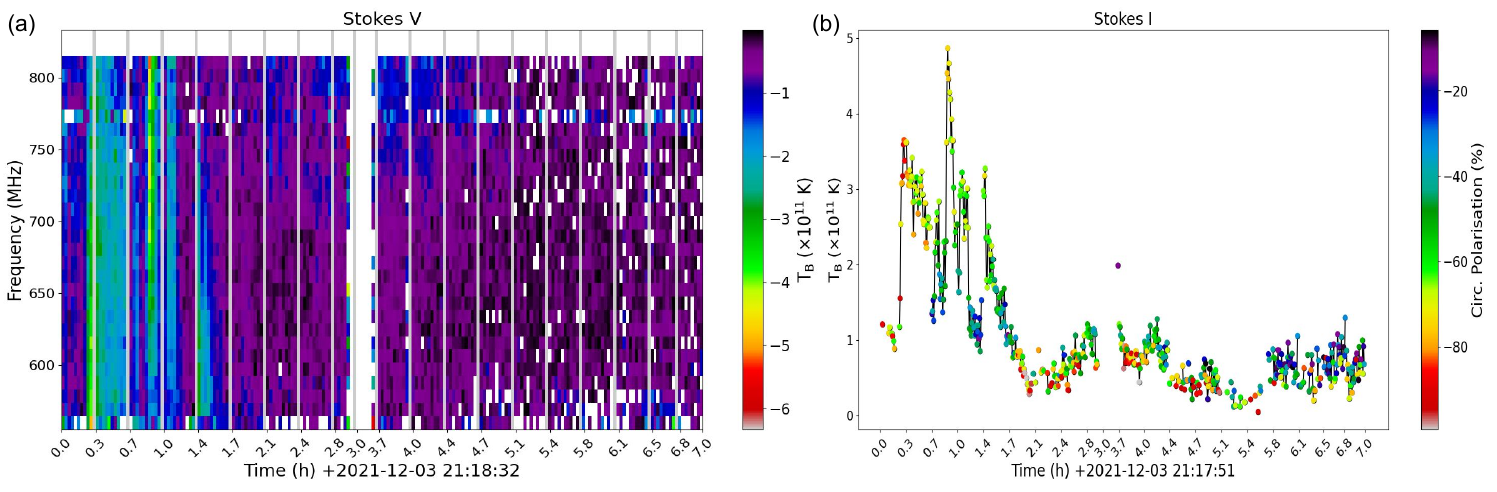}
    \vspace{-0.8cm}
    \caption{(a): STOKES V VISAD for AD\,Leo (M4V) at 50\,s and 5\,MHz resolution. (b) Band averaged STOKES I lightcurve with circular polarisation \% marked (Mohan et al., in prep).}
    \label{fig3:VISAD}
\vspace{-0.2cm}
\end{figure}

Unlike solar observations, observations of the cool main-sequence stars do not resolve the star. So imaging is not strictly necessary to explore spectro-temporal evolution of stellar activity.
However, a time averaged imaging of the stellar field is mandatory to obtain a model for the background sky across spectral channels. This background model visibilities can then be subtracted from the corrected data to end up with purely stellar visibility data.
VISibility Averaged Dynamic spectrum (VISAD) routine works on this purely stellar spectro-temporal visibility data.
VISAD centres the visibility data to the expected location of the star and computes the mean visibility as a function of frequency and time. The routine lets the user choose the required frequency and time averaging so as to detect stellar emission in the dynamic spectral plane.
VISAD can generate dynamic spectrum in all STOKES parameters and polarisation supported by the data.
It can also generate band averaged light curves.
Figure~\ref{fig3:VISAD} shows STOKES V dynamic spectrum and band-averaged circular polarisation lightcurve for an active M-dwarf AD Leo, made with VISAD applied on uGMRT data (Mohan et al., in prep).
The code is equipped with parallel processing capabilities and uses some of  the functionalities of Common Astronomy Software Applications~\citep[CASA; ][]{casa} 
A version of this routine has already been released via Github\footnote{\MYhref{https://github.com/atul3790/Visibility-averaged-DS}{https://github.com/atul3790/Visibility-averaged-DS}}

The following sections will present some of the key science cases explored using the high resolution snapshot spectroscopic data, applying tools like SPREDS and VISAD.

\section{Solar activity}
\label{sec:sun}
Sensitive modern interferometers let us explore the ubiquitous weak radio bursts across all four axes of variability (time, frequency ($\approx$ height), angular sky coordinates) at much finer resolution than previously possible~\citep[e.g.][]{div11,Atul17,kontar2017}.
Tools like SPREDS has been a used to study the dynamics of accelerated particles, propagation of waves, instabilities and turbulence across corona at active and quiet solar regions~\citep{Div22_SKAO_india}.
Some of the interesting discoveries and research avenues that the modern data and tools have opened up are the following.
\paragraph{\textbf{Solar flares: }}Solar flares often generate a variety of radio and mm bursts. The solar radio bursts, especially the strong ones, have been studied over decades and classified into different burst types\citep[see,][for an overview]{wild1970,SolarRadio_book,white2007_radiobursts}. 
With the advent of modern sensitive interferometers and spectrographs we are now able to detect weak events, with energy ranges similar to and weaker than micro-flares\citep[e.g.][]{Ramesh13_picoflares,  suresh17_waveletMWA, Rohit18_weakest0.2SFUflare, mondal21_WINQES_UVcounterpart} Besides, the high spectro-temporal cadence of these instruments also let us probe the fine structures and fast quasi-periodic variability in the emission\citep[e.g.][]{mugundhan2017, kontar2017,sharykin2018_LOFAR_dnn_withtypIIIb, Atul19_QPO, Mohan2021b}. Some aspects of these new discoveries pertaining to weak radio bursts will be covered in the paragraphs below.

\paragraph{\textbf{3-D Quasi-periodic pulsations (QPP): }}
%The regions of weak coronal activity on the sun are usually associated with events which often barely perturbs the ambient physical fields rather than causing a significant change as in the case of eruptions, massive outflows or strong flares.
Weak radio bursts often trigger quasi-periodic pulsations (QPPs) in the ambient physical fields observable in the image plane. Earlier radio studies had identified such fast $\sim$\,s timescale QPPs during radio bursts in the full disk integrated flux dynamic spectra. 
However, the evolution of these QPP sources, their spatial distribution and structure remained unexplored until recently, due to limitations in high fidelity spectroscopic snapshot imaging.
Recent snapshot spectroscopic imaging studies of the radio counterparts of microflare sources, using SPREDS, revealed that the intensity QPPs are often associated with correlated QPPs in source sizes and sky orientation~\citep{Atul19_QPO,Atul19_ARTB_microflare} giving them a three dimesionality (size, orientation and flux density).
\cite{atul21_structQPPs} showed that these 3-D QPPs had 2 different modes of variability - `S' (size -flux density anti-correlated evolution) and `T' (size - orientation correlated evolution). These modes systematically evolved in tandem with the thermal energy and the local magnetic structure during the microflare.
This result reveal a novel means to use QPP diagnostics to explore the local magnetic field evolution at weakly flaring regions, which are otherwise difficult to probe.

\paragraph{\textbf{Coronal turbulence and waves:}} 
At coronal heights beyond $\sim$4\Rsun and into the solar wind regime,  interplanetary scintillation techniques had been used along side other radar-based techniques to infer the density turbulence~\citep{coles1991,manoharan1990,anantharamaiah1994_VLA_angBroad,sasi2017}. 
However, in the inner corona we lacked a robust means to characterise the local turbulence.
Since coherent radio emission originate at frequencies close to \nup, which is related to local \N, wave propagation is heavily influenced by the strength of local density fluctuations (\dnn)~\citep[see,][]{steinberg1971,robinson_scat1983,thejappa1998,Arzner1999,Kontar19_Arznercopy}.
This causes rapid ($\sim$\,s) size and shape variability in the observed burst sources alongside observed intensity. Wideband snapshot spectroscopic images lets us explore this variability across inner coronal heights, using techniques like SPREDS~\citep{kontar2017,Atul19_QPO}. 
Employing the physical model proposed by \cite{Arzner1999}, \cite{Atul19_QPO} derived \dnn\ across inner coronal heights (1.4 - 1.45\,\Rsun) using MWA data.
Later \citep{Atul21_dNN_vsht} extended the study to bursts observed across the entire MWA operational bandwidth (80 - 240\,MHz) and estimated the turbulence characteristics of solar corona within $\sim$ 1.4 - 1.8\,\Rsun. 
Similarly by modelling the observed radio wave scattering effects at frequencies below 80 MHz, turbulence characteristics have been explored up to $\sim$\,2.2\,\Rsun\ by several authors~\citep[e.g.][]{mugundhan2017,sharykin2018_LOFAR_dnn_withtypIIIb,Chen20_snapshottypeIIIEvol_turb}. 

Radiowave propagation across corona is also influenced by local plasma wave modes. 
These effects can be used to infer the spatial scales of the disturbances and properties of their progenitors. 
For instance, type-III striae bursts offer a means to explore propagating Langmuir waves~\citep[e.g.][]{Thejappa2018LangmuirSI,Reid21_typeIIIb_langmuirwave}.
\paragraph{\textbf{Radio - CMEs: }}
CMEs are the most violent eruptions happening on the Sun which leave significant impacts on space weather. 
However, the particle acceleration sites and their evolution during CMEs remain less understood. 
High cadence spectro-polarimetric imaging studies have started revealing interesting details on this aspect~\citep[see,][for an overview]{Carley20_CMEImagingreview,Chen23_CME_SSImging_modeling}.
CMEs often produce the type-II radio bursts which are believed to be associated with particle acceleration sites around CME shocks\citep{Nelson85_typeII_CMEshockaccl}. 
The high time resolution radio imaging explorations of these sources using instruments like LOFAR help explore the dynamical evolution of particle acceleration sites across CME shock fronts~\citep[e.g.][]{Morosan19CME_shockreg}. Combined with high energy data from multiple spacecrafts at different vantage points, a 3D reconstruction of the evolution of the CME shock and the associated particle acceleration regions can be done~\citep[e.g.][]{zucca18_3DCME_shockprop,2022A&A...668A..15M}.

Apart from exploring particle acceleration sites, radio observations of split-band type-II bursts offer a means to compute the Mach number of CME shocks. Combined with white light observations they can help estimate the mean magnetic field strength at the shock regions\citep{Anshu17_splitbandtypeII_B,2018JASTP.172...75M}.
Polarimetric studies (degree of circular polarisation) of CME associated radio bursts can also provide estimates of the mean magnetic field strengths at CME shocks~\citep[e.g.][]{2014ApJ...796...56S,2017ApJ...843...10K,2022ApJ...940...80R}.
However, a more direct spatially resolved magnetic field estimation at regions associated with the CME  can be performed using imaging observations and subsequent modeling of the gyro-synchrotron emission from various regions in the image plane~\citep[e.g.][]{Bastian01_CMEImg_NRH, Maia07_radioCMEimg, Bain14_TypeIVM_img}.
Such studies however remained difficult due to lower sensitivity of previous generation instruments.
Sensitive new generation instruments like MWA are now able to image CME shock fronts that produce order of magnitude weaker radio flux than previous detections~\citep{mondal20_CMEgyro,Debo23_polCME}. Modelling the polarised gyro-synchrotron radio spectra from CMEs, the local magnetic field estimates have also been made~\citep{Debo23_polCME}.

\paragraph{\textbf{Quiet sun: }}
{ Quiet Sun, unlike the name suggest, is expected to be teeming with weaker energy release events which collectively dump enough energy to maintain its high background temperature\citep[see,][ for an overview]{Parker63_nanoflare,Klimchuk06_CorHEating_overview,klimchuk15_CorHeat_keyaspects}. 
Several X-ray to radio band studies have shown evidence for high emission variability at quiet regions or during relatively quiet periods of solar activity\citep[e.g.][]{Mercier97_nanoflaresNoisestorm, Pauluhn07_NanoflareSUMERdata,Testa13_nanoflares_in_Moss,Testa14_Moss_nanoflare_Scipaper,Viall17_nanoflares_inARs}. 
High dynamic range snapshot spectroscopy imaging facilities have enabled the exploration of emission variability across regions of interest on the quiet sun disk.
For instance, studies using high sensitivity MWA data have revealed ubiquitous non-thermal activity down to $\sim$\ 0.2 SFU, approaching sub-picoflare energy levels for the first time~\citep[e.g.][]{suresh17_waveletMWA, Rohit18_weakest0.2SFUflare}.  
These are about an order of magnitude weaker than previously reported weakest flaring event\citep{Ramesh13_picoflares}}. With high resolution snapshot spectroscopic imaging, certain authors have also explored such weak flares across the solar disk and demonstrated its ubiquitous nature~\citep{mondal21_WINQES_UVcounterpart,Rohit22_QS_picoflares}. 
Besides, data from MWA and LOFAR have also helped explore the radiowave scattering/propagation effects across quiescent solar disk~\citep{Rahman19_quietCorhole,rohit20_QS_propagationeffect,Peijin22_LOFAR_QSmaps}.
Recent advancements in imaging polarimetry have lead to direct detection of quiet sun magnetic field strengths, which are otherwise impossible to obtain at 1.1 - 2\Rsun heights in the corona~\citep{Patrick19_Polcal_QSmaps,Debo23_MWAPolcal}. 
All these observational aspects are crucial in improving the existing solar coronal emission models like FORWARD and making more reliable estimates of the true source energy levels, spatial distribution and morphology ~\citep{Gibson16_FORWARD,rohit20_QS_propagationeffect}. 

\subsection{Chromospheric tomography with ALMA}
ALMA opened up a new era of solar chromospheric exploration with its high sensitivity and angular resolution at sub-arcsecond scale (depending on the chosen array configuration and frequency). ALMA can provide both total power (TP) brightness temperature maps of the solar disk at $\sim$\,8\,s cadence~\citep{White17_ALMA_TPmaps} and also zoomed in interferometric images of selected portions on the solar disk at $\sim$\,2\,s resolution~\citep[see,][for an overview on data and results]{Vasco22_SALSA}.
The TP maps provide an overall idea of the mm emission characteristics across the disk and a means to study the disk averaged sun-as-a-star spectral evolution~\citep{Atul22_EMISSAII}.
Meanwhile, the zoomed in multi-band interferometric imaging datasets at high spatio-temporal resolution let us explore propagating waves, shocks and varying chromospheric structure across height~\citep[e.g.][]{Sven16_ALMA_science,2021RSPTA.37900185E,shahin21_overview_ALMAB3B6Oscill, Nindos22_ChromDyn_mmBand,Nancy22_ALMAOscill}.
Besides, \cite{Juan22_structQPOs} reported structural QPPs akin to the metric 3-D QPPs in high resolution ALMA Band 3\,(100\,GHz) movies. 

\section{Stellar activity}
%Cool stars which host the most number of Earth-like exoplanets\citep{Bashi20_occurrence_of_smallplanetsFGK} in their habitable zones also happen to be the ones with active outer atmospheres that drive flares, eruptions and other high energy phenomena, which can jeopardise the stability of the atmospheres of nearby planets in habitable zones \citep{Vidotto13_highB_forclosebyplanets}.
Sensitive spectroscopic snapshot imaging capability has lead to significant advancements in the field of stellar activity.
The advancement is two fold; one in the form of extending the source `detectability' horizon out to distances as high as $\sim$ 300\,pc~\citep{2020A&A...639L...7V, callingham21_dM_LOTSS,vedantham22_xray-radio_rel,Callingham22_vLOTSS}, while the other in detecting fine scale spectro-temporal variability~\citep{Osten2008,2018ApJ...862..113C}.
These studies have already questioned our understandings about radio flare flux levels, and their relation with other waveband fluxes and physical properties~\citep[e.g.][]{vedantham21_mechCohEmissStars,vedantham22_xray-radio_rel}.
New instruments have also lead to discoveries of star-planet interaction signatures~\citep{vedantham21_SPI}.
Tools like VISAD have helped to identify weak flaring/star-planet interaction signatures in long-duration wideband monitoring data~\citep[e.g.][]{Osten2008,villadsen19_Cohbursts_but_notypeII}.

\subsection{Chromospheric tomography - a tool to characterise stellar activity}
%Ground and space based instruments provide plenty of data primarily in X-ray to optical wavebands \citep[e.g.][]{GaiaDR32021,HARPS2014_datarelease} and also in radio bands using facilities like the JVLA, ATCA etc. \citep[e.g.][]{Villadsen14_First_detect_SLS_inRadio_VLA, Zic20_typeIV_ProximaCen}. 
Based on the Xray to optical and radio band observables, different stellar activity indicators have been constructed and studied as functions of physical parameters like stellar mass (\Ms), \teff, age, rotation period (\Prot), magnetic field strength etc.
Common activity indicators include, the ratio of Ca-II H-K flux to bolometric flux (\RHK, \cite{Noyes84_RHK}) and the X-ray to bolometric flux ratio (\Rx).
Though these indicators provide qualitative trends between activity and various physical parameters, a quantitative characterisation remains difficult due to the large variability in their values resulting from their complex dependencies on multiple parameters and physical processes (thermal, non-thermal and propagation effects)~\cite{stepien94_Defn_Rx+Ro_Vs_activity_n_manyCorCurves,Pace13_RHK_Drastic_variability}.
This issue limits the insights that can be gained regarding the emergence of different levels and nature of activity from the atmospheres of different stellar types~\citep{Atul22_EMISSAII}.

\subsubsection{Exploring robust indicators of Chromospheric heating/activity}
\begin{figure}[t]
    \centering
    \includegraphics[width=\textwidth]{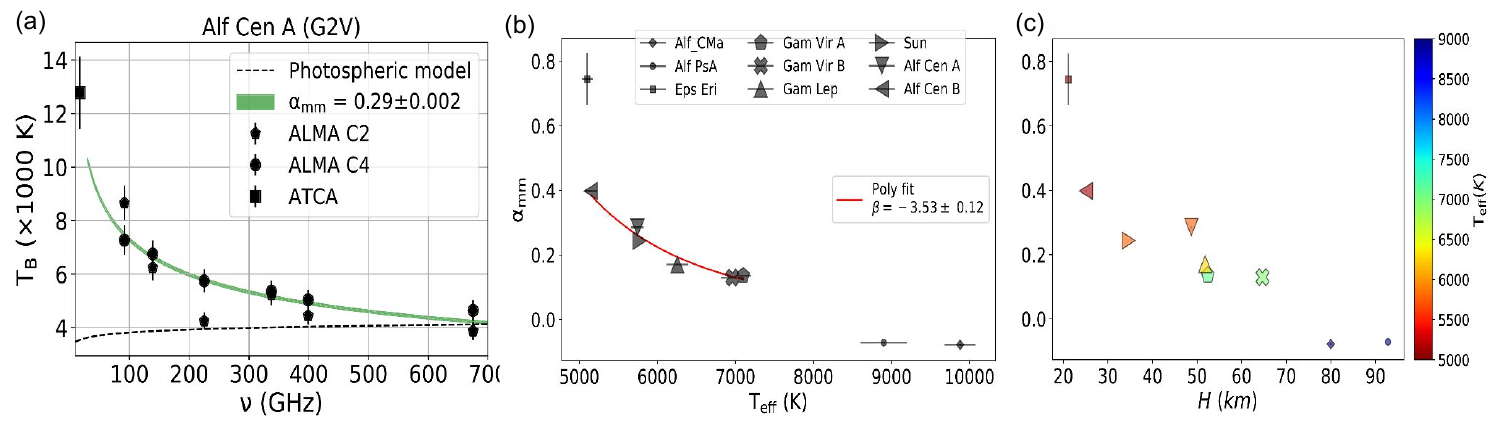}
    \vspace{-0.8cm}
    \caption{(a): mm-\Tbp\ for $\alpha$\,Cen\,A. Dotted line shows the photospheric emission model and the green curve shows the best fit powerlaw to \Tbp\ with index \alfmm. Different markers show data from different observation cycles (denoted by `C') and telescopes. (b): \alfmm\ versus \teff\ for the ALMA detected stellar sample~\citep{Atul22_EMISSAII}. Bigger markers denoted old ($>$1\,Gyr) stars. Red line shows the power-law fit. A-type stars (\teff\,$>$ 8000\,K) are shown for comparison. (c): \alfmm\ versus pressure scale height for the same sample}
    \label{fig4:alfmm}
\vspace{-0.2cm}
\end{figure}
Observations and related models suggest that the atmospheric structure of cool stars significantly vary across spectral type or equivalently \teff~ \citep[e.g.][]{Donati09_Rev_Bfield,Linsky16_Stellar_chromRev}. 
%The chromosphere, acts as an important layer between the photosphere and corona, playing a crucial role in the transport of energy and mass during both quiescent and active periods.\citep[see,  ][for a review]{Linsky16_Stellar_chromRev}.
For the cool active stars it is known that the atmospheric heights extending from chromospheres to corona show signs of steady heating driven by their strong magnetic activity, which is also the driver of different space weather  phenomena~\citep{Linsky16_Stellar_chromRev,Vidotto14_B_Vs_age_n_rot}.  
It is therefore highly desirable to construct a new and more reliable observational indicator of 
%We need an observational means to characterise 
the quiescent atmospheric thermal structure of cool stars, which is closely linked to the atmospheric magnetic activity.

Recent solar and stellar observations demonstrated the unique tomographic potential of the mm-brightness spectrum to deduce the chromospheric thermal structure in cool stars (Fig.~\ref{fig4:alfmm}a) \citep[e.g.][]{2017SoPh..292...88W,White20_MESAS,2018MNRAS.481..217T,Atul21_EMISSAI}.
Owing to their sensitivity down to $\sim$ 20$\mu$Jy with a few hours integration, NOEMA and ALMA enable the detection of a sample of nearby stars in the mm band out to $\sim$25\,pc.
Using ALMA detected sample of cool stars, \cite{Atul22_EMISSAII} showed that mm-\Tbp\ spectral indices (\alfmm; \Tbp$\propto \nu^{\mathrm{-\alpha_{mm}}}$) were positive for F -- M dwarfs (\teff$\sim$7200 -- 3000\,K). 
%shown to track the atmospheric heating gradients in cool stars~\citep{Atul22_EMISSA_activityindex}. 
Since lower frequencies probe higher heights, a positive \alfmm\ indicates progressively hotter and active outer atmospheres. 
The A-type stars with \teff$>$9000\,K showed negative \alfmm\ as expected from stars with no upper photospheric heating. 
The \alfmm\ versus \teff\ showed an inverse trend suggesting that A6-9 type ( \teff$\sim$ 7500 - 8000\,K) stars could be the ones close to \alfmm$\sim$0, marking the point of rise of chromospheric heating and activity in the main-sequence (Fig.~\ref{fig4:alfmm}b).
Spectral line studies also suggest the same about late A-type stars~\citep{2002ApJ...579..800S, Linsky16_Stellar_chromRev}.
Besides, \cite{Atul22_EMISSAII} demonstrated that \alfmm\ can be a robust activity indicator since the power-law correlation functions between \alfmm\ and stellar physical parameters have much lower uncertainty ranges than those obtained with \RHK\ and \Rx.
Also, \alfmm\ relates inversely to pressure scale heights in the atmosphere, as expected from a proxy to atmospheric thermal gradients (Fig.~\ref{fig4:alfmm}c).

%However, the sample of ~\cite{Atul22_EMISSAII} is incomplete to draw statistically significant conclusions on the physical dependencies of chromospheric heating gradients probed indirectly by \alfmm.
%There were 6 old stars ($>$1Gyr) with just one young star. So all correlation functions were derived only for old stars. Among the old stars, there was no sufficient spread in the values of ages and rotation rates, which are as crucial physical determinants of activity.

\section{Conclusions and outlook}
\label{sec:conclusions}
The new generation sensitive, mm-radio interferometers are bringing about revolutionary changes in the exploration of solar/stellar activity.
These instruments collectively provide extremely wideband data from 10\,MHz to 1\,THz at $\lesssim$\,s resolution.
This snapshot spectroscopic data let us perform a tomographic exploration of solar/stellar atmospheres since different frequencies are sensitive to different heights in the atmosphere, ranging from upper photosphere to outer corona.
The modern interferometers also employ a compact `large-N' architecture which ensures dense u-v sampling leading to massive improvement in imaging fidelity and dynamic range for fine temporal ($\lesssim$ 1\,s) and spectral averaging ($\sim$ 10\,kHz resolution in metrewave band; $\sim$ 2\,GHz (continuum) and 0.01 - 10\,MHz (spectral line) in mm band).
This lets us do a continuous sampling of active atmospheric heights.

Modern instrumentation and data analysis tools/pipelines have lead to several novel discoveries and kick started new research avenues in solar/stellar activity in the recent years. However, much of these results and ventures are in an early phase, requiring more data and better models with detailed physics.
There is hence a strong need to monitor the sun and stars, and gather more data during different periods of solar/stellar activity, varying in local physical conditions, energy levels and space weather impact. 
Analysing such datasets hold a lot of discovery potential to identify novel phenomena like the examples mentioned in the article, and also will help construct better physical models.
Large volume of datasets is also an essential precursor to a detailed statistical characterisation/classification of the different types of active phenomena and exploring their space weather impacts/significance. 

\textit{Acknowledgements:} This work is supported by the Research Council of Norway
through the EMISSA project (project number 286853) and the Centres of Excellence scheme, project number 262622 (``Rosseland Centre for Solar Physics''). This research made
use of NASA's Astrophysics Data System (ADS). AM is grateful to the developers of Python3 and various packages namely Numpy \citep{numpy}, Astropy \citep{astropy}, Scipy \citep{scipy} and Matplotlib \citep{matplotlib}.

%%%DO NOT REMOVE THIS LINE otherwise the bibtex bibliography won't work anymore %%%
%%%%%%%%%%%%%%%%
\newcommand{\newblock}{}
\bibliographystyle{mnras}
%%%%%%%%%%%%%%%%
\bibliography{EMISSA_allref.bib}

\end{document}